\documentclass[onecolumn,preprint,pre,showpacs,preprintnumbers,amsmath,amssymb,11pt]{revtex4}

\def\b{\mathbf}\def\Re{{\rm Re}}
\def\e{\epsilon}
\def\bs{{\boldsymbol{\sigma}}}

\usepackage{graphicx}
\usepackage{dcolumn}
\usepackage{bm}
\usepackage{color}

\begin{document}

\title{Stokesian jellyfish: Viscous locomotion of  bilayer vesicles}

\author{Arthur A. Evans$^1$}
\email{aevans@physics.ucsd.edu}
\author{Saverio E. Spagnolie$^2$}
\author{Eric Lauga$^2$}
\email{elauga@ucsd.edu}
\affiliation{$^1$ Department of Physics, University of California San Diego, 9500 Gilman Drive, La Jolla CA 92093-0354}
\affiliation{$^2$ Department of Mechanical and Aerospace Engineering, University of California San Diego, 9500 Gilman Drive, La Jolla CA 92093-0411.}
\date{\today}

\begin{abstract}

Motivated by recent advances in vesicle engineering, we consider theoretically the locomotion of  shape-changing bilayer vesicles at low Reynolds number. By modulating their volume and membrane composition, the vesicles can be made to change shape quasi-statically in thermal equilibrium. When the control parameters are tuned appropriately to yield periodic shape changes which are not time-reversible, the result is a net swimming motion over one cycle of shape deformation. For two classical vesicle models (spontaneous curvature and bilayer coupling), we  determine numerically the sequence of vesicle shapes through an enthalpy minimization, as well as the fluid-body interactions by solving a boundary integral formulation of the Stokes equations. For both models, net locomotion can be obtained either by continuously modulating fore-aft asymmetric vesicle shapes, or by crossing a continuous shape-transition region and alternating between fore-aft asymmetric and fore-aft symmetric shapes. The obtained hydrodynamic efficiencies are similar to that of other low Reynolds number biological swimmers, and suggest that shape-changing vesicles might provide an alternative to flagella-based synthetic microswimmers. 

\end{abstract}

\maketitle

\section{Introduction}

The preeminence of viscous dissipation over inertial effects at low Reynolds numbers leads to many interesting consequences for life and engineering efforts at the micron-scale. In particular, swimming at zero Reynolds number is impossible using time-reversible motions, a result known as the Scallop theorem \cite{Purcell:1977p29}. As a result, at least two actuation degrees of freedom are necessary to generate locomotion. The breaking of this time-reversal symmetry has been studied both from a mathematical point of view, and in the context of modeling real organisms \cite{Brennen:1977p30,cd04,Lauga:2007PF,Fu:2008p518,Lauga:2009p421}. Unlike in high Reynolds number flows, such as those relevant in describing the swimming of fish and flying of birds, fluid motion at low Reynolds numbers is set almost instantaneously by the time-dependent geometries of the immersed bodies. Thus it is natural to inquire about the shapes of immersed (and possibly fluctuating) cell membranes, and their relationships to locomotion.

Membranes composed of lipid bilayers are ubiquitous in nature, and the study of bilayer vesicles as a model system for  biological cells has yielded significant insight into their  behavior \cite{Seifert:1997p475,Kas:1991p647}.  In addition to the biological relevance of lipid bilayer vesicles, or liposomes, advances in self-assembly have paved the way for other types of vesicles to be developed experimentally \cite{Antonietti:2003p627,Li:2009p629}.  Vesicles assembled from block copolymers \cite {Kukula:2002p626}, liquid crystal amphiphiles \cite{Yang:2006p624}, and membranes with embedded proteins or anchored polymers \cite{Lipowsky:1995p628,Wang:2005p623,Guo:2009p630,Breidenich:2009p631,Antunes:2009p622} all have tunable material properties which can be manipulated with unprecedented control \cite{Dobereiner:2009p218,Lee:1999p477}.  It is also well known that many biological cells actively modify or maintain the shapes of their membranes \cite{McMahon:2005p201,Veksler:2007p136}, either for developmental \cite{Huang:2006p115} or locomotive processes \cite{Grimm:2003p648,Bottino:2002p649}. 

Recently, synthetic microswimmers inspired by the locomotion of eukaryotic cells have been successfully designed in experiments~\cite{dreyfus05_nature}, exploiting the planar beating of a flagellum-like organelle. Beyond biomimetic engineering, other small-scale synthetic swimmers or swimming strategies have also been proposed, both theoretically and experimentally \cite{Purcell:1977p29,becker03,najafi04,kts05,lk07,leoni08,leshansky08,Lauga:2009p421,Spagnolie09}. One recently-studied example is a self-propelled colloidal particle which exploits asymmetrically-distributed chemical reactions to swim in a viscous fluid~\cite{howse07,golestanian07}. 

In the same spirit, we consider theoretically in this paper a novel swimming mechanism based on prescribed shape transformations of a bilayer vesicle. By modulating only its volume and membrane composition, the vesicle can be made to change shape quasi-statically in thermal equilibrium. For two different theoretical  vesicle models, we determine numerically the vesicle shapes through an  enthalpy minimization, and the fluid-body interactions by solving a boundary integral formulation of the Stokes equations. When the control parameters are tuned appropriately to yield periodic but not time-reversible shape changes, we show that 
net locomotion can be obtained. Swimming arises either by continuously modulating fore-aft asymmetric vesicle shapes, or by crossing a continuous shape-transition region and alternating between fore-aft asymmetric and fore-aft symmetric shapes. In addition, the calculated  hydrodynamic efficiencies are shown to be similar to that of other common low Reynolds number propulsive mechanisms. 

Our paper is organized as follows. We begin with a general discussion of the practical realization of  controlled shape-changing vesicles, in particular the relevant time scales, and the possible actuation mechanisms. Two classical curvature-mediated vesicle models (spontaneous curvature and bilayer coupling) are presented, and the formulations used for the shape calculation and the numerical fluid-interaction model are introduced. We then discuss examples of vesicle shape cycles that yield a swimming motion, examine the fluid flow that develops around the vesicles during their deformation cycles, and compute the corresponding swimming speeds and hydrodynamic efficiencies.

\section{A Roadmap to Vesicle Locomotion}

A vesicle immersed in a viscous fluid experiences a highly coupled array of forces, such as those generated by membrane tension, internal pressure, membrane (bending) elasticity, and the surrounding viscous fluid dynamics. In the microscopic, viscous environments relevant to our consideration, the Reynolds number, $\Re$, is very small: $\Re=\rho\,U_cL_c/\mu \ll 1$, where $\rho$ is the fluid density, $\mu$ is the fluid shear viscosity, and $U_c$ and $L_c$ are characteristic velocity and length scales of the vesicle.  The fluid behavior at low Reynolds number is highly dependent upon the immersed boundary geometry, and the resultant forces include not only local, but also non-local responses to its motion. 

A general study of vesicle dynamics should take non-equilibrium shapes into account, as even simple liposomes that can be created in situ can interact relatively quickly with the environment.  It is possible to design experiments where carefully constructed initial conditions and lipid species lead to equilibrated vesicle shapes that are non-trivial, but in order to apply morphological changes and induce locomotion, a reversible parameter-changing mechanism is desirable.  

For our first approach to vesicle swimming, we consider in this paper a ``stiff membrane'' regime.  The characteristic time of membrane relaxation in a viscous fluid is given by $t_{rel}=\mu r_0^3/\kappa$, where $\kappa$ is the elastic bending modulus of the membrane, and $r_0$ is a characteristic radius of curvature.  If we choose the maximum radius of the vesicle for the characteristic length scale $L_c$, then $r_0\lesssim L_c$.  For parameter variation significantly slower than the membrane relaxation rate, {\it i.e.} for a cycle time scale $t_{cycle}\gg t_{rel}$, then we operate safely within the decoupled regime. In this case, we may thus assume that there are no hydrodynamically induced shape changes, and that the shapes are determined quasi-statically in equilibrium.  Using this time scale $t_{rel}$, we can also set a maximum swimming velocity scale, $U_c=\kappa/\mu r_0^2$.  Similar scaling arguments have been made in Refs.~\cite{Lipowsky:1999p114,Brochard:1975p632}.   For biologically relevant systems in water, $\kappa\approx 100$ k$_B$T, $\mu\approx 10^{-3} $Pa \,s, $\rho=1$ g/cm$^3$, and $r_0\lesssim  1-10$ $\mu$m, leading to $L_c\approx 1-10$ $\mu$m,  $U_c\approx 1-10$ $\mu$m/s, $t_{rel}\approx0.01-1s$ and $Re\approx10^{-4}$.  For a vesicle with length scale $L_c=10$ $\mu m$, diffusive time scales are approximately $10^4 s$, and thus negligible for the time being.  In addition, we neglect thermal fluctuations in the determination of the vesicle shape, as they  come in as a perturbation about the mean equilibrium shape of order $(k_BT/\kappa)^2$, which is very small under most conditions \cite{Seifert:1997p475}.

There are a number of different physical means by which a vesicle shape can be changed in a controlled fashion, and the methods could be  different depending on the type of vesicle considered.  We will consider two such means, internal volume changes and local membrane compositional changes. 

One experimentally feasible example of a possible volume-changing mechanism is a light-induced osmotic change. In an ordinary biological membrane the bilayer is embedded with numerous proteins, many of which are sensitive to mechanical forces, chemical gradients, or light. The protein bacteriorhodopsin, for example, is sensitive to green light, and in response to a signal the protein opens and closes like a valve \cite{Karp99}. The presence of such ion channels or active proteins on the surface of a membrane can cause osmotic changes of the fluid volume contained within the vesicle \cite{Lee:1999p477}.  Recently, vesicle volume control was demonstrated via pH modulation of block copolymer networks along the surface of membrane \cite{Yu:2009p611}.  The vesicles in this study were well separated from regimes associated with morphological transition, and thus changes in osmotic pressure induced only a volume change, leading to a ``breathing" vesicle. 

Adjusting the membrane composition requires a more indirect experimental approach. Some bilayers are composed of different species of constituent parts, leading to an inherent mismatch between the intrinsic curvatures. In other words, there is an intrinsic curvature that would develop across the bilayer in the absence of other considerations.  Because of the inherent difficulty in measuring these quantities it is likely to be more difficult to specify an exact change from one value of intrinsic curvature to another.  
However, the actual process of changing the intrinsic curvature can be achieved through inducing chemical changes of the lipid constituents of the membrane \cite{Yu:2009p611}, or by conformational changes of polymers grafted to the surface of the vesicle \cite{Nikolov:2007p577}.

\begin{figure}[t!]
\includegraphics[width=3in]{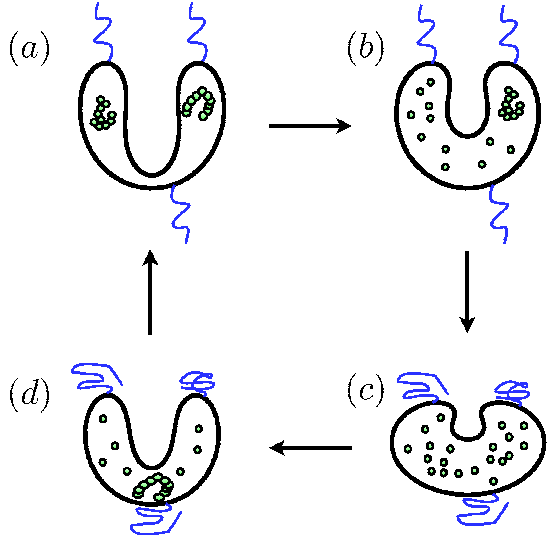}
\caption{\label{schematic} Schematic illustration of a possible control mechanism for vesicle shape-change and swimming: Axisymmetric bilayer vesicle with embedded reactive polymers and polymers grafted to its surface. 
$(a)\rightarrow(b)$:  Short frequency light impinges on the vesicle, catalyzing a de-polymerization reaction amidst the particle chains, and increasing the fluid volume available to the vesicle. 
$(b)\rightarrow(c)$: A second frequency of light induces the grafted polymers to coil up, inducing an entropic repulsion from the membrane and changing the macroscopic morphology. 
$(c)\rightarrow(d)$: The dispersed particles begin to polymerize back to their initial configuration, deflating the vesicle. 
$(d)\rightarrow(a)$: A third frequency of light is used to uncoil the polymers, relaxing the entropically induced curvature and returning the vesicle to its initial state.}
\end{figure}

By combining two shape-changing mechanisms, it would in theory be possible to achieve a periodic shape cycle which is not time-reversible, yielding a net locomotion. One of many possible configurations that could produce a cycle in shape space is displayed schematically in Fig.~\ref{schematic}, where we consider a bilayer vesicle with embedded reactive polymers and with polymers grafted to its surface. In the first step (Fig.~\ref{schematic}a$\to$b), a photo-chemical polymerization reaction is catalyzed by green (short wavelength) light, and the polymer chains in the interior of the vesicle disperse into a solution of particles, thus increasing the available volume within the vesicle.  At a later time, another frequency of light (red, or longer wavelength) impinges on the vesicle, and the grafted polymers change from a distended to a coiled conformation, inducing an entropic repulsion and changing the curvature of the membrane (Fig.~\ref{schematic}b$\to$c).  Over time the dispersed particles will polymerize and return the vesicle to its original volume (Fig.~\ref{schematic}c$\to$d), and finally a third frequency of light (blue or very short wavelength) can be used to change the conformation of the polymers to distended once more, returning the vesicle to its original state (Fig.~\ref{schematic}d$\to$a).  

While osmotic volume change or chemical-induced composition alteration are two possible experimental methods, not only these examples in no way constitute the full set of possibilities, but also they might be difficult to implement experimentally. Other experimental techniques already exist (see Refs.~\cite{Yu:2009p611,Lee:1999p477}), or may be developed in the near future that could be more suited for controlled two-parameter change.

Rather than suggest specific experimental methodology whose specifics would depend not only on the particular material of the bilayer vesicles, but also on the parameter alternation methods, we adopt in this paper a simplified modeling approach  that highlights the qualitative pieces that are required in order to transform a motionless vesicle into a locomotive cargo-carrier. In parallel to the the various practical mechanisms that could be used to implement such shape changes experimentally, it is  of fundamental interest to ask theoretically the question of prediction and performance. Would shape change indeed lead to locomotion of the vesicle? How efficient would it be? Can we quantitatively predict the resulting swimming speed and the work done against the fluid to achieve it? This is the approach taken in this paper.  Considering two simplified vesicle models, and for slow modulations of the vesicle shapes, we introduce below a computational framework able to quantitatively predict swimming kinematics and performance.

\section{Dynamics of Coupled Fluid-Body System}

\subsection{Vesicle physics}

While real biological membranes have multiple constituents, all interacting in non-trivial ways, minimal models can still help to illuminate the fundamental physics of such systems.  For length scales on which a membrane is approximately flat a Monge parameterization can be employed \cite{Ramaswamy:2000p165,Reigada:2005p75,Campelo:2007p473}, but for a closed bilayer vesicle the curvatures can become very large and the small geometric gradient assumption may break down. In order to characterize the shapes of such objects, an  enthalpy must be extremized and the full nonlinear shape equations so generated must be solved.  There are many models that could be used to describe the physics of curvature-mediated vesicle morphology. In this paper we will consider two classical models as case studies. These formulations, known respectively as the spontaneous curvature and bilayer coupling models, have both been used in classical work \cite{Seifert:1991p430} and correspond to different interaction dynamics between the membrane monolayers.  Both of these models also include exactly two free parameters, which enable us to explore the breaking of the Scallop theorem, and the generation of locomotion via a change in morphology. 

The enthalpy functional, $F$, in the spontaneous curvature model takes the following form \cite{Seifert:1991p430}
\begin{gather}
F=\frac{\kappa}{2}\int_{S(t)}{\left(C_1+C_2-C_0\right)^2dS}+\Sigma \,A +P\,V,\label{FreeEnergy}
\end{gather}
where $C_1$ and $C_2$ are the principal membrane curvatures, and $\Sigma$ and $P$ are Lagrange multipliers which constrain the surface area $A$ and volume $V$ (physically they correspond to the membrane tension and pressure difference across the interface). In Eq.~\eqref{FreeEnergy}, $S(t)$ denotes the time-dependent surface boundary, and  $C_0$ is the spontaneous curvature, which introduces an inherent mismatch in equilibrium preference of the membrane curvature. This quantity along with a fixed volume and surface area completely specifies the ensemble. Thus the spontaneous curvature model has area, volume, and integrated spontaneous curvature constrained, and we select as the control parameters the volume $V$ and the spontaneous curvature $C_0$ (the fixed surface area merely selects the overall size of the vesicle). 

In contrast, in the bilayer coupling model, the enthalpy functional $G$  assumes the area difference $\Delta A$ between the membrane monolayers to be constant.  One possible representation of this area difference is in terms of the integrated mean curvature, 
\begin{gather}
M=\displaystyle\int_{S(t)}{(C_1+C_2)}\,dS.
\end{gather}
Then the area difference is $\Delta A= 2 h M+O(h^2/A)$, where $h$ is the distance between monolayers \cite{Seifert:1991p430}.  The enthalpy then takes the form
\begin{gather}
G=\frac{\kappa}{2}\int_{S(t)}{\left(C_1+C_2\right)^2dS}+\Sigma' \,A +P\,V+Q\,M,\label{FreeEnergy2}
\end{gather}
where $\Sigma'$, $P$, and $Q$ are Lagrange multipliers associated with $A$ (area), $V$ (volume) and $M$ (integrated mean curvature) respectively. We select as control parameters the volume $V$ and the integrated mean curvature $M$.

It is important to note that the functionals $F$ and $G$ are related via a Legendre transform, $\left(\Sigma',Q\right)\rightarrow\left(\Sigma+\kappa \,C_0^2/2,-2\kappa \,C_0\right)$, and thus describe the same system in a different ensemble.  Physically, the spontaneous curvature model corresponds to a bilayer in which the monolayer admits stretching or compression during bending, and thus finds an equilibrium distribution that has a preferred curvature. If the bilayer is composed of more than one species of lipid, each of which has a different preferred curvature ({\it i.e.} radius of gyration), it is likely that the membrane will actually prefer to be in a non-flat state.  Conversely, the bilayer coupling model corresponds to a system that enforces that both monolayers are incompressible. The area difference between monolayers  stays approximately constant on the timescales relevant to our consideration, and as long as the distance between layers remains very small this implies that the integrated mean curvature also remains constant.

\begin{figure}[t!]
\includegraphics[width=2.6in]{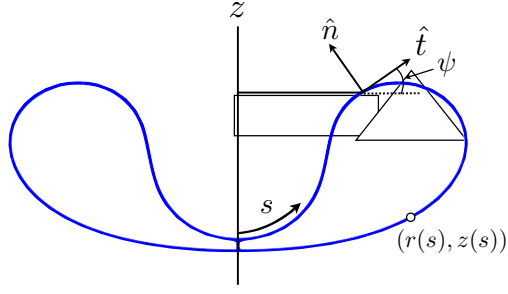}
\caption{\label{shape} Parameterization of an axisymmetric bilayer vesicle.  We assume axisymmetry about the z-axis. The surface is described by $\b{x}=\left(r(s,t),z(s,t)\right)$ in cylindrical coordinates, with $s$ an arc-length parameter, $\b{\hat{t}}$ the unit tangent vector, $\b{\hat{n}}$ the outward pointing normal vector, and $\psi$ the angle between the x-axis and $\b{\hat{t}}$.}
\end{figure}

\subsection{Determination of the vesicle shape}

Assuming an axisymmetric vesicle shape, the body surface $S(t)$ is parameterized at each time $t$ as illustrated in Fig.~\ref{shape}. The arc-length measured along the surface in the $x$-$z$ plane is denoted by $s\,\in[0,L]$, with $\b{\hat{t}}$ the unit tangent vector, $\b{\hat{n}}$ the outward pointing normal vector, and $\psi$ the angle between the x-axis and $\b{\hat{t}}$. The body surface is represented in cylindrical polar coordinates,
\begin{gather}
\b{x}(s,\phi,t)=\b{\tilde{x}}(s,\phi,t)+z_0(t)\b{\hat{z}}=(r(s,t)\cos(\phi),r(s,t)\sin(\phi),z(s,t)+z_0(t)),
\end{gather}
where $\phi\in[0,2\pi)$ is the azimuthal angle, the surface $\b{\tilde{x}}$ is taken to have its center of volume at the origin, and $z_0(t)$ is a translation of that center of volume which depends upon the fluid interaction. Under this parameterization, the principal membrane curvatures are $C_1=\partial \psi/\partial s$ and $C_2=\sin{\psi}/r$. Upon insertion into either of the enthalpy functionals $F$ or $G$, and performing a variational extremization, we obtain the following system of first-order ordinary differential equations to describe the energetically stationary vesicle shapes at time $t$ \cite{Seifert:1991p430}
\begin{gather}
\label{shape_eqns0}
\psi_s=K,\\
K_s=-\frac{K}{r}\cos\psi + \frac{\gamma}{r}\sin\psi +\frac{\cos\psi\sin\psi}{r^2} +\frac{1}{2}Pr\cos\psi,\\
\gamma_s=\frac{(K-C_0)^2}{2}-\frac{\sin^2\psi}{2r^2} + Pr\sin\psi +\Sigma, \\
\label{shape_eqns}
r_s=\cos\psi.
\end{gather}
Here $K$ is an auxiliary function used to make the system of equations first-order (physically it corresponds to the curvature), $\gamma$ is the Lagrange multiplier that enforces the interdependence of $\psi$ and $r$, and the subscript s denotes a derivative with respect to the arclength. The vesicle shape at time $t$ is set by Eqs.~(\ref{shape_eqns0}-\ref{shape_eqns}), subject to the four boundary conditions $r(0,t)=r(L,t)=\psi(0,t)=0$ and $\psi(L,t)=\pi$. Once the angle $\psi$ is determined from the above, $z(s,t)$ is set by an integration of $z_s=\sin(\psi)$, where the constant of integration is chosen such that the center of volume of the surface $\b{\tilde{x}}$ is at the origin. The vertical position $z_0(t)$ has no bearing on the vesicle shape determination, and we hold off further discussion on its dynamics until the following section.

For the spontaneous curvature model, constraints on the unknown integration length $L$, the surface area $A$, the volume $V$, and the two constant Lagrange multipliers $P$ and $\Sigma$ are imposed as
\begin{gather}
A_s=2\pi r,\,\,\,\,V_s=\pi r^2 \sin\psi,\,\,\,\,P_s= 0, \\
\Sigma_s= 0,\,\,\,\,L_s=0 \label{constraint_eqns}.
\end{gather}
Defining $R_0$ as the radius of the sphere with surface area $A$, the boundary conditions for the five constraint equations above are $A(0)=V(0)=0$, $A(L)=4\pi R_0^2$, $V(L)=4\pi R_0^3 v/3$, where $v$ is a dimensionless ``reduced volume.'' Due to the Lagrange function being independent of the arc-length $s$, the ``Hamiltonian" is a conserved quantity and we have $\gamma(0)=0$ (see Refs.~\cite{Seifert:1991p430,Jiang:2007p259}). Also defining a reduced spontaneous curvature $c_0=C_0R_0$, we finally obtain the vesicle morphology as set by the two parameters $(v,c_0)$.

In the bilayer coupling model, Eqs.~(\ref{shape_eqns0}-\ref{constraint_eqns}) are solved with two additional constraints. First, the integrated mean curvature $M$ is controlled, $M_s=\pi(rK+\sin{\psi})$, and second,  a new Lagrangian constraint enters, $Q_s=0$.  The system is now closed with boundary conditions on the integrated mean curvature: $M(0)=0$ and $M(L)=4\pi R_0\Delta a$, where $\Delta a$ is the reduced surface area difference between monolayers, $\Delta a=\Delta A/8\pi R_0h$.  In this case the vesicle morphology is set by the two parameters $(v,\Delta a)$, and the reduced spontaneous curvature $c_0$ has been removed from the shape equations via the Legendre transform given above.

Equations~(\ref{shape_eqns0}-\ref{constraint_eqns}) are solved numerically. Due to coordinate singularities in the derivatives of r and z at the poles, the shape is determined on the contracted interval $s\in[L\,\delta,L\,(1-\delta)]$ for $(L\,\delta)\ll1$, and Taylor-expanded versions of the boundary conditions are applied. For example,
\begin{align}
\nonumber r(L\,\delta,t)&=r(0,t)+(L \,\delta)\, r_s(0,t)+O\left((L\delta)^2\right)=(L\,\delta)\,r_s(L\,\delta,t)+O\left((L\delta)^2\right)\\
&=(L\,\delta)\cos(\psi(L\,\delta))+O\left((L\delta)^2\right)\approx L\,\delta.
\end{align}
To compute the shapes using either model, the arc-length is discretized using $m$ uniformly spaced grid points, $s_i$, with $s_1=L\delta$ and $s_m=L(1-\delta)$. A collocation method is then applied in a formulation and implementation similar to that recently used by Jiang et al. \cite{Jiang:2007p259}.  We employ a standard continuation scheme in order to interpolate solutions from one point in the parameter space $(v,c_0)$ or $(v,\Delta a)$ to neighboring points.

By extremizing the enthalpies $F$ or $G$, the shape equations give only stationary solutions, not necessarily the lowest energy solutions.  A numerically determined shape may correspond to an energy saddle point, maximum, or minimum. Although it is possible that the lowest energy state may not be achievable for a non-equilibrium shape change, for our purposes we will examine the minimum energy shapes, and thus a ``phase diagram" for the possible shapes is of great use.  Just as in a more conventional phase transition, shape transformations correspond to transitions between different symmetry states.  Since we consider only axisymmetric shapes here, spherical solutions have the highest symmetry state. For small perturbations around spherical shapes, the solution can be represented as
\begin{gather}
r(s,t)=R_0\left(1+\sum_{\ell=0}^\infty B_{\ell 0}Y_\ell^0(\theta(s),\phi=0,t)\right),
\end{gather}
where the functions $Y_\ell^0$ are the spherical harmonics, and the constants $B_{\ell 0}$ can generate symmetry breaking.  Because we consider only axisymmetric vesicles, only the $m=0$ spherical harmonics (of the $Y_\ell^m$) contribute to the sum, and the angle $\theta$ is given by $\tan{\theta}=r/z$.  While it is not possible to produce an analytical solution using this formulation, it is useful for understanding the morphological transitions in terms of symmetry breaking. For example, breaking $\ell=2$ symmetry $(B_{2 0} \neq 0)$ leads to a prolate or oblate shape, while breaking $\ell>2$ symmetry can give more complicated shapes, such as the so-called ``pear" or ``stomatocyte'' shapes \cite{Seifert:1997p475}. In our numerical investigation, symmetry is frequently exploited in order to efficiently compute the equilibrium shape. In regions of multiple stability the solution branches that correspond to lowest energy shapes must be chosen, and by inserting numerically an initial symmetry breaking the algorithm used can more readily converge upon the appropriate solution.

\subsection{Fluid-body interaction}

Modulation of the dimensionless parameter set $(v,c_0)$ or $(v,\Delta a)$ generates quasi-static deformations which in turn lead to motion in the surrounding fluid medium. Given that the Reynolds number is small, the dynamics of the fluid surrounding the vesicle is effectively governed by viscous dissipation and is well modeled by the incompressible Stokes equations,
\begin{gather}
\nabla \cdot \bs = 0,\,\,\,\,\nabla \cdot \b{u}=0,
\label{StokesEqs}
\end{gather}
where $\bs=-p\b{I}+2\mu \b{E}$ is the Newtonian stress tensor
with $p$ the pressure, $\b{u}$ the fluid velocity, and $\b{E}$ the
symmetric rate-of-strain tensor, $\b{E}=\frac{1}{2}(\nabla\b{u}+(\nabla\b{u})^T)$. The fluid equations are made dimensionless by scaling velocities upon $U_c$, lengths upon $L_c$, and time upon $t_{rel}=L_c/U_c$. Since the surface area $A=4\pi R_0^2$ is constant, we define the characteristic length scale by this radius, {\it i.e.} $L_c=R_0$.  Henceforth, the swimming velocity is understood to be dimensionless, and each shape cycle occurs over a unit in dimensionless time. 

A no-slip condition is applied on the body surface.  For a given path through the parameter space $(v,c_0)$ or $(v,\Delta a)$, the resulting sequence of instantaneously determined shapes set uniquely the ``surface deformation velocity'' $\b{u_d(x},t)$; namely,
\begin{gather}
\b{u_d}(\b{x}(s,\phi,t),t)=\frac{\partial \b{\tilde{x}}}{\partial t}(s,\phi,t).
\end{gather}
In addition, the surface moves as a rigid body along the $\b{\hat{z}}$ direction due to axisymmetry, with velocity $\b{U}=U\b{\hat{z}}=z_0'(t)\b{\hat{z}}$. The no-slip condition is thus written as $\b{u}(\b{x},t)=U\b{\hat{z}}+\b{u_d(x},t)$. 

To close the system of equations describing the fluid-body interaction, we assume that no external forces are acting upon the vesicle, and thus
force and torque balance give
\begin{gather}
\int_{S(t)}\bs(\b{\b{x})\cdot \hat{n}(\b{x})}\,dS =
0,\,\,\,\,\,\int_{S(t)}\b{x} \times[\b{\bs(\b{x})\cdot
\hat{n}(\b{x})]}\,dS = 0 .
\label{ForceBalance}
\end{gather}

The computation of the swimming velocity is performed using a standard double-layer boundary integral formulation of the Stokes equations. The details of this formulation and numerical method are presented in the appendix.

In addition to computing the swimming velocity, we consider a possibly more important quantity, the hydrodynamic efficiency. This swimming efficiency is defined as (see Ref.~\cite{Childress81})
\begin{gather}
\eta_{H}=\frac{\Big\langle\b{U}\cdot \b{F}\Big\rangle}{\Big\langle\displaystyle\int_{S(t)}{\left(\b{U+u_d}\right)\cdot \b{f}} \, dS\Big\rangle}=\frac{\Big\langle\b{U}\cdot \b{F}\Big\rangle}{\Big\langle\displaystyle\int_{S(t)}{\b{u_d}\cdot \b{f}}\, dS\Big\rangle},\label{etaH}
\end{gather}
where $\b{f=-\bs \cdot \hat{n}}$ is the force density acting on the fluid at the body surface, $\langle \cdot \rangle$ denotes a time-average over a full shape cycle, and $\b{F}=6\pi\mu\, a\, U\b{\hat{z}}$ is the force required to move a sphere of radius $a$ at a speed $U$. At each time we use the maximum vesicle radius, $a(t)=\|r(s,t)\|_\infty$. The first term in the denominator of Eq.~\eqref{etaH} integrates to zero due to the zero-net force condition (Eq.~\eqref{ForceBalance}).  The computation of the fluid stress $\bs$ is significantly more involved than the computation of the swimming velocity. We employ a numerical method for computing $\bs$ based on the evaluation of a hypersingular integral which may be derived from the double-layer formulation of the fluid velocity. The framework and numerical approach are described in the appendix, and a more detailed description of the method and examples of its use will be featured in a subsequent paper.

Physically, $\eta_{H}$ measures the proportion of work done by the vesicle against the surrounding fluid which is used for swimming purposes, and is typically on the order of 1\% for biological cells. Note that the swimming efficiency only measures the hydrodynamic efficiency, not a total efficiency. For example, the bending energy of the vesicle is not captured in this measure. The inclusion of bending costs into swimming efficiency measures has recently been proposed to study optimal locomotion strategies in flagellated cells, but presents an avenue of inquiry beyond the scope of this paper \cite{sl10}.

\section{Vesicle locomotion by shape-change}

As stated in the introduction, due to the linearity and time-reversibility of Eqs.~\eqref{StokesEqs}, any time-reversible geometrical surface deformations cannot result in a net locomotion. This result is known as the Scallop theorem, in reference  to the sole, time-reversible motions available to a small scallop (opening and closing) \cite{Purcell:1977p29}. As a consequence of this constraint, a single degree of freedom is insufficient for swimming. Two degrees of freedom are however sufficient to generate a swimming motion, as first described in Ref.~\cite{Purcell:1977p29}, and as we shall show presently for the systems of interest.

\subsection{Spontaneous curvature model}

\begin{figure}[t!]
\includegraphics[width=5in]{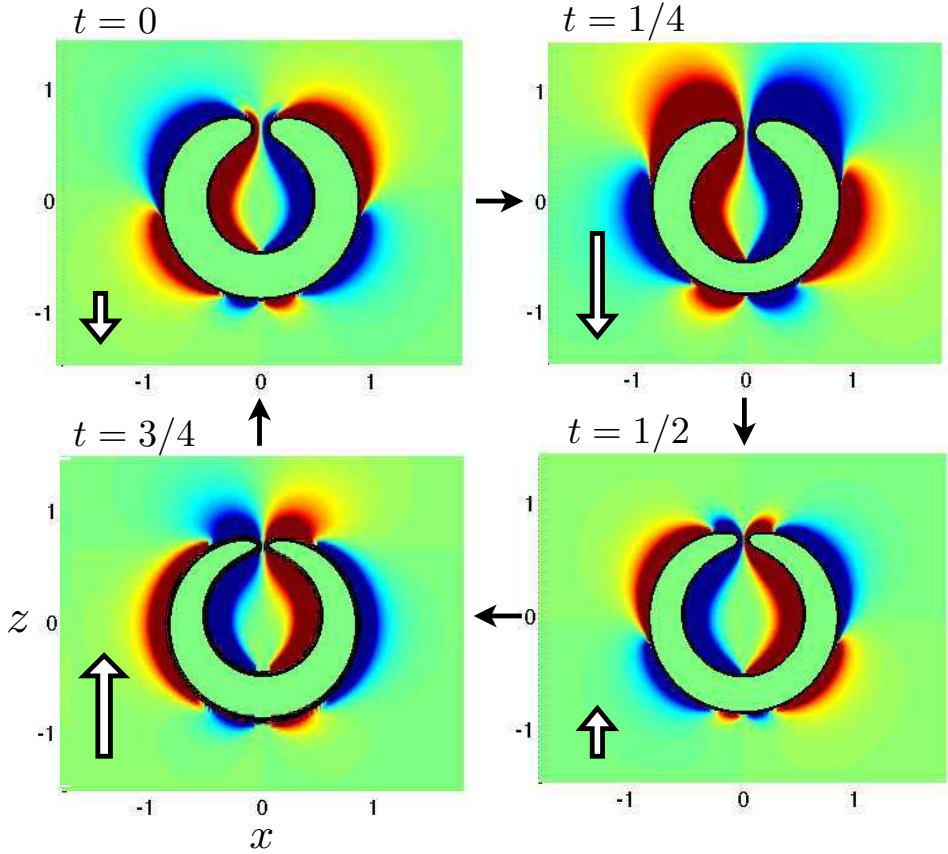}
\caption{\label{flowstom} (color online) Stomatocyte shapes and vorticity profiles produced using the spontaneous curvature model, with $v(t)=0.425+0.125 \cos(2\pi t)$, $c_0(t)=-0.1+0.3 \sin(2 \pi t)$. Positive vorticity, corresponding to counter-clockwise rotation, is shown in red, and negative vorticity, corresponding to clockwise rotation, is shown in blue. Hollow arrows indicate the instantaneous swimming velocity. During one cycle, the vesicle experience net locomotion in the $-\b{\hat{z}}$ direction.}
\end{figure}

We begin by presenting a characteristic shape cycle that can be generated by adjusting the reduced volume and spontaneous curvature, $(v,c_0)$, in a periodic fashion. By selecting a specific elliptical path in the $(v,c_0)$ parameter space, namely $v(t)=0.425+0.125 \cos(2\pi t)$, $c_0(t)=-0.1+0.3 \sin(2 \pi t)$, the resulting shape cycle is not time-reversible; hence, the constraints of the Scallop theorem are bypassed, and locomotion may be achieved.  For these parameters the vesicle shapes are always stomatocytes, and the neck separating the internal sphere of fluid from the external fluid is very small. Figure~\ref{flowstom} shows the corresponding minimal energy vesicle shapes at four times, along with the vorticity generated in the surrounding fluid by the body deformation, $\boldsymbol{\omega}=\nabla \times \b{u}$. Positive vorticity, corresponding to counter-clockwise rotation, is shown in red, and negative vorticity, corresponding to clockwise rotation, is shown in blue. Hollow arrows indicate the instantaneous swimming velocity of the vesicle, while the plain arrows indicate the direction of time. At zero Reynolds number the swimming velocity, external flow, and swimming efficiency are determined uniquely by the time-dependent surface geometry and surface deformation velocity, so we need not consider the internal flow dynamics (which may in general depend upon the means of modulating the parameters $(v,c_0)$).

From $t=0$ to $t=1/4$ the vesicle volume is decreasing while the spontaneous curvature is increasing.  The decrease in volume draws fluid into the stomatocyte cavity, while the surface material near the opening to the cavity moves inward nearly tangentially to the surface itself. While the deformation velocity is normal to the surface near the north and south poles ($s=0$ and $s=\pi$), the deformations are elsewhere primarily tangential, and vorticity is created as the fluid is sheared accordingly. At $t=1/2$ the vesicle volume is minimal, and the fluid volume inside the stomatocyte cavity is beginning to decrease.  From $t=1/2$ to $t=3/4$, the vesicle volume increases while the spontaneous curvature continues to decrease to its minimum value.  This can best be understood by observing that when $c_0<0$ the membrane prefers a total negative curvature, and as can be seen at $t=3/4$, the internal cavity of the vesicle takes its smallest value, maximizing negative curvature.  The increasing volume expels fluid from the cavity, and leads to a reversing of the sign of the vorticity.  The overall sequence of asymmetric shapes  is not time-reversible, leading to a net swimming velocity taking place in the $-\b{\hat{z}}$ direction.

\begin{figure}[t!]
\includegraphics[width=6.4in]{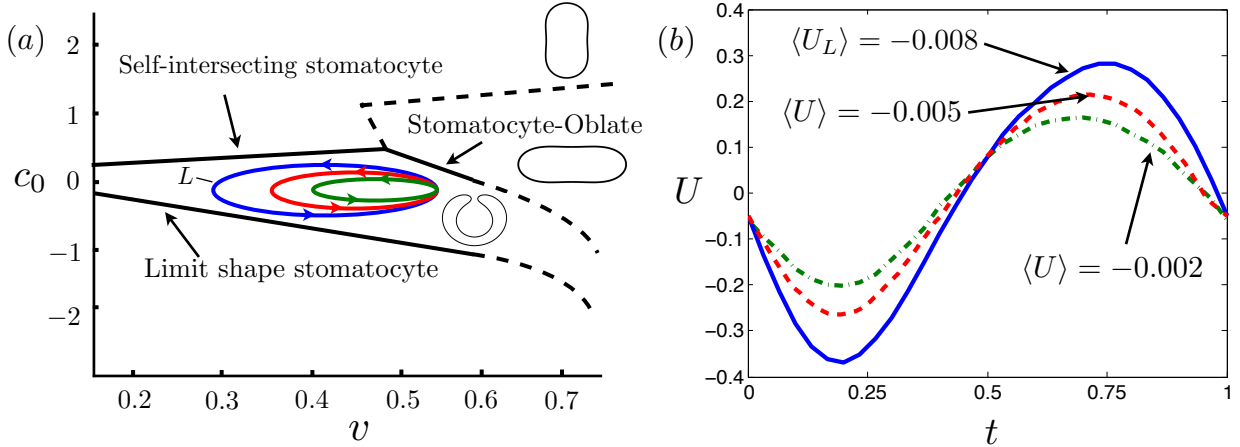}
\caption{\label{swim} (color online) (a) Phase diagram for the spontaneous curvature model in the $(v,c_0)$ parameter space.  Solid lines are our numerically-calculated lines that denote morphological transitions, while the dashed lines are qualitative, and adapted from Ref.~\cite{Seifert:1991p430}.
(b) Three velocity profiles, corresponding to the elliptic paths through parameter space indicated in (a), with the largest velocities achieved along the elliptic path enclosing the greatest area.}
\end{figure}

A phase diagram for the minimal energy shapes using the spontaneous curvature model is presented in Fig.~\ref{swim}a. The limit lines correspond to discontinuous morphological transitions, and therefore cannot be crossed in our quasi-static shape-change approach. One critical line corresponds to vesicles whose north and south poles self-intersect, and a second line corresponds to stomatocyte shapes that have a vanishing opening between the external fluid and the cavity within ({\it i.e.} the shapes are two spheres, one contained entirely within the other).  A third line marks the discontinuous phase transition between stomatocyte and oblate shapes. More details may be found in Ref.~\cite{Seifert:1991p430}. 

Beyond the symmetry constraints imposed by the Scallop theorem, other symmetry breaking is necessary in order for a body to achieve a net motion from a periodic shape cycle. Namely, the body surface must express fore-aft asymmetry in order to swim preferentially in any direction. Hence, parameter paths in the regions of phase space corresponding to prolate or oblate vesicle shapes cannot yield a net motion. However, paths which correspond to stomatocyte or pear shapes are fore/aft asymmetric and can swim.  Since the area in phase space that contains pear shapes is very small, we will only examine the swimming stomatocytes. The largest elliptic path shown in Fig.~\ref{swim}a corresponds to the shape cycle shown in Fig~\ref{flowstom}. The associated time-evolution of the vesicle  center of mass velocity is  shown in Fig.~\ref{swim}b, along with two other velocities corresponding to elliptic paths enclosing smaller areas in Fig.~\ref{swim}a. 

We see in Fig.~\ref{swim} that the larger the area of the cycle in parameter space, the faster the vesicle swims. In fact,  
the mean velocity roughly  scales as the square-root of the area enclosed by the elliptic path  of phase space.
Drawing on an analogy with thermodynamics, cycles with larger area in the appropriate ensemble space do more work, and thus we might expect that the transduction of shape deformation into mechanical work would exhibit similar behavior.  Although our equivalent to an equation of state is  too complicated to show a simple relationship between swimming velocity and the area enclosed in this  phase space, the basic idea appears to remain valid.

We finally note that the net translation during each shape cycle in each case is small compared to the amplitude of the motion, and even smaller when compared to the maximum vesicle radius. The swimming velocities and hydrodynamic efficiencies of shape cycles in the spontaneous curvature model are also small.  The maximum velocity achieved for the cycles shown is $\langle U \rangle = -0.008$, while we calculate an efficiency of $\eta_H= 0.4\%$. 

\subsection{Bilayer coupling model}

\begin{figure}[t!]
\includegraphics[width=6.4in]{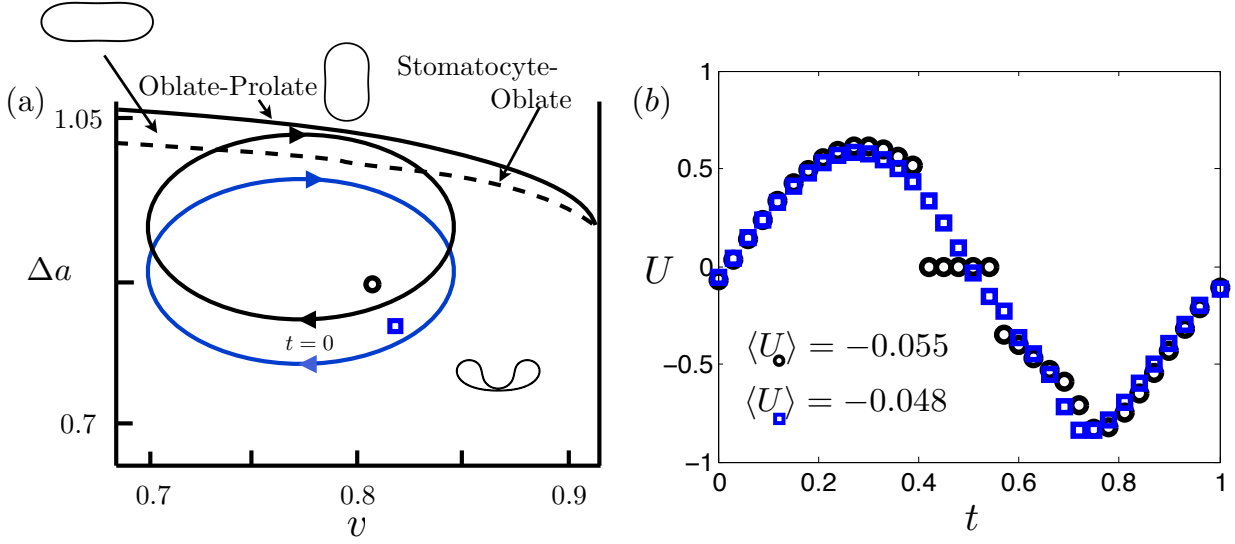}
\caption{\label{bcmodel} (color online) (a) Phase diagram for the bilayer coupling model in the $(v,\Delta a)$ parameter space, adapted from Ref.~\cite{Seifert:1991p430}. The dashed line indicates a continuous transition, while the solid line indicates a limit shape.  Both lines are shown  schematically in order to exaggerate the difference between the shape cycles.  The two elliptical cycles considered  enclose the same area in phase space, but one crosses the transition line. (b) Swimming velocity of the vesicle as a function of time, for the two shape cycles shown in (a). The squares denote the continuously varying velocity of the lower cycle in (a), which is similar to what we observed for the spontaneous curvature model. The circles correspond to the upper cycle in (a) and involves a shape transition, and there is a portion of the cycle during which the vesicle has zero swimming velocity due to fore-aft symmetry.}
\end{figure}

We now consider the bilayer coupling model, for which a schematic phase diagram is shown in Fig.~\ref{bcmodel}a.  Although in the spontaneous curvature model there are no continuous transitions between oblate and stomatocyte shapes, the interesting feature of the bilayer coupling model is the presence  of a continuous stomatocyte-oblate transition.  The upper (solid) line in Fig.~\ref{bcmodel}a denotes a limit line between oblate and prolate shapes, while the lower (dashed) line represents a continuous transition between stomatocyte and oblate shapes.

In order to examine how breaking or restoring oblate ($\ell=2$) symmetry relates to swimming, we now consider two shape cycles with equal enclosed area in phase space, as shown in Fig.~\ref{bcmodel}a. The upper cycle crosses the continuous transition line, while the lower cycle remains in the stomatocyte region. 

\begin{figure}[t!]
\includegraphics[width=5in]{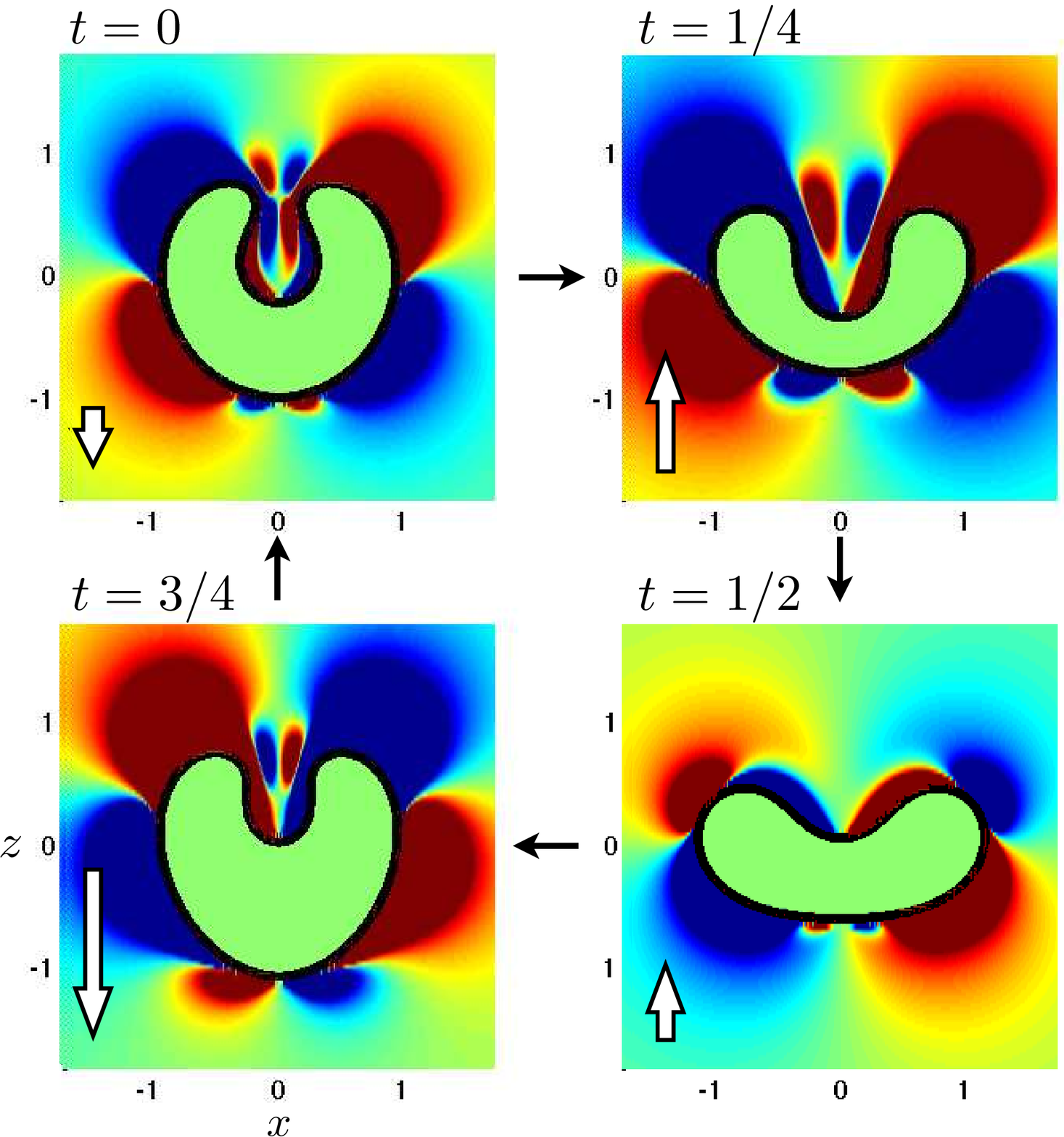}
\caption{\label{changing}  (color online)  Vesicle shape cycle using the bilayer coupling model, with $v(t)=0.775+0.075\sin(2\pi t)$, $\Delta a(t)=-0.14\cos(2\pi t)+0.89$, corresponding to the 
lower cycle of Fig.~\ref{bcmodel}a.  This vesicle does not change morphological symmetry states during the swimming cycle and remains within the stomatocyte domain.  Hollow arrows denote the instantaneous swimming velocity.}
\end{figure}

The vesicle shapes in the lower cycle of Fig.~\ref{bcmodel}a are displayed in  Fig.~\ref{changing}. They correspond to a modulation of the  volume and surface area difference between monolayers for the vesicle as  $v(t)=0.775+0.075\sin(2\pi t)$, $\Delta a(t)=-0.14\cos(2\pi t)+0.86$. From $t=0$ to $t=1/4$ the vesicle volume is increasing, expelling fluid from the cavity and pushing fluid away from the surface of the membrane. Due to the larger amount of surface area facing the aft end of the vesicle, the net motion during this quarter-cycle is forward.  From $t=1/4$ to $t=1/2$, the ``lobes" of the vesicle move downwards, propelling the vesicle upwards, albeit at a decreasing rate.  This portion of the motion resembles the characteristic undulatory shape of a jellyfish, albeit one at zero Reynolds number.  Between $t=1/2$ and $t=3/4$, the vesicle deflates and the lobes begin to move upwards again, with the material points of the lobes moving almost completely tangentially to the surface. This creates a vortex dipole at the lobes, leading to the stagnation point that can be seen in the figure.  Finally, in the last quarter cycle, the vesicle encloses itself and returns to the starting position.  We calculate a mean swimming velocity of $\langle U \rangle=-0.048$,  and a hydrodynamic efficiency of $\eta_H= 0.6\%$.

\begin{figure}[t!]
\includegraphics[width=5in]{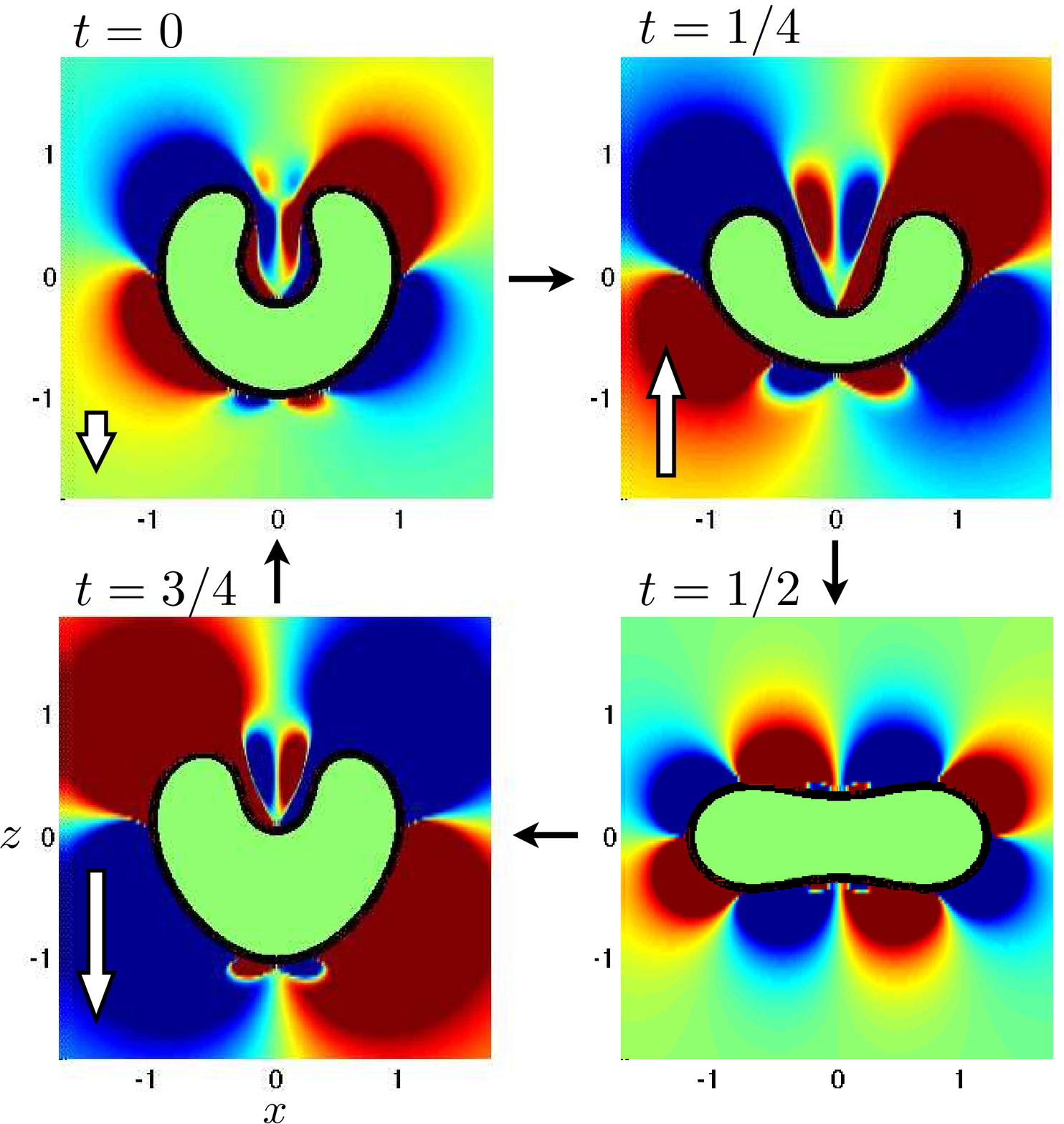}
\caption{\label{changing2}  (color online) Vesicle shape cycle using the bilayer coupling model across a continuous phase transition, with $v(t)=0.775-0.075\sin(2\pi t)$, $\Delta a(t)=0.14\cos(2\pi t)+0.89$, and  corresponding to the upper cycle of Fig.~\ref{bcmodel}a.  This vesicle is oblate for a small part of the cycle, precluding swimming  by symmetry, but a net locomotion occurs over the entire cycle.  Hollow arrows denote instantaneous swimming velocity.}
\end{figure}

The upper elliptical cycle of Fig.~\ref{bcmodel}a,  with shapes illustrated in  Fig.~\ref{changing2},  follows the parameter path $v(t)=0.775+0.075\sin(2\pi t)$, $\Delta a(t)=-0.14\cos(2\pi t)+0.89$, which lies above the continuous stomatocyte-oblate phase transition line from $t\approx 0.45$ to $t\approx 0.55$. During this portion of the cycle the vesicle has exactly  zero swimming velocity due to the fore/aft symmetry of oblate shapes. Between $t=0$ and $t=1/4$, the volume and area difference are decreasing, leading the nearly oblate shape into a clearly stomatocyte configuration.  For our purposes, we will not address the spontaneous symmetry breaking that is associated with crossing a transition line, but simply assume that once broken, the cycle will break the symmetry in the same way during each cycle.  In the example shown, the stomatocyte inflates as it assumes a more oblate shape, expelling fluid from the cavity and producing vorticity along the lobes.  As the vesicle continues to deflate from $t=1/4$ to $t=1/2$, the lobes sweep downwards, moving the stomatocyte upwards as it assumes a perfectly oblate shape.  At $t\approx 0.42$ the shape transitions into an oblate shape, precluding any net swimming by symmetry. The swimming velocity as a function of time is shown in Fig.~\ref{bcmodel}b.  As the oblate vesicle deflates, at $t\approx 0.58$ the stomatocyte symmetry state is entered once more, the lobes sweep upwards, and the vesicle moves downwards. Despite the presence of a becalmed period during the vesicle does not move, the cycle that involves the shape transition yields a larger mean velocity than the lower cycle, $\langle U \rangle =-0.055$, and an increased hydrodynamic efficiency, $\eta_{H}=0.7\%$.

As previously noted, crossing the shape transition line between stomatocyte and oblate shapes indicated in Fig.~\ref{bcmodel}a yields a continuous shape change. However, if we exploit the analogy with phase transitions, we note that some quantities must be discontinuous across the transition.  Without exploring the details of a dynamic phase transition in the context of vesicle locomotion, although the order parameter is continuous, derivatives of the order parameter need not be so.  In other words, the material at a given point $s$ along the boundary experiences a continuous positional change, but a discontinuous velocity relative to the  center of mass of the body as the parameters are varied continuously through the transition line. The discontinuous relative material velocity then generates the discontinuous swimming velocity seen in Fig.~\ref{bcmodel}b for the body which exhibits the oblate shapes for part of its periodic cycle.

Interestingly, even though the area enclosed in phase space by the two cycles illustrated in Fig.~\ref{bcmodel}a is the same, the relationship between parameter space, efficiency, and swimming velocity is not evident.  The upper cycle shown in Fig.~\ref{changing2} has a larger mean swimming speed and is more efficient than the cycle shown Fig.~\ref{changing}, suggesting that the vesicle can increase its efficiency by passing through a phase transition.

\section{Discussion}

In this paper, we have shown computationally that it is possible for a bilayer vesicle to swim under a prescribed shape change using two different vesicle models.  By modulating the vesicle volume and either its preferred curvature (spontaneous curvature model) or the surface area difference between membrane monolayers (bilayer coupling model), the vesicle can be made to undergo deformations which are not time-reversible, yielding therefore a net swimming motion. Net locomotion can be obtained either by continuously modulating  fore-aft asymmetric vesicle shapes (stomatocytes), or by crossing a continuous shape-transition region with fore-aft symmetric shapes, and alternating therefore between  fore-aft asymmetric and fore-aft symmetric shapes.

At first sight, the swimming efficiencies obtained in this paper appear to be low. For the swimming stomatocyte shown in Fig.~\ref{swim}, the efficiency is on the order of $0.4\%$, while for the bilayer coupling model we calculate an efficiency of $0.6\%$ for a non-transitioning vesicle, and $ 0.7\%$ for a vesicle that undergoes a transition from stomatocyte to oblate.  However, it is known from many theoretical studies that the  hydrodynamic efficiency of swimming  microorganisms, such as flagellated bacteria or spermatozoa, is on the order of 1 to $ 2 \%$ (see Ref.~\cite{Lauga:2009p421} and references therein).  Our results indicated therefore that the equilibrium morphologies of bilayer vesicles, together with their appropriate modulations as is done in this paper, lead to locomotion means which are almost as efficient as those displayed by biological cells, and  might therefore provide an interesting alternative to flagella-based synthetic micro-swimmers. Further optimization of the size and shape of cycle in parameter space will likely lead to swimming vesicle outperforming the efficiency of flagellated cells. In addition, a swimming vesicle has the advantage that the swimmer and the cargo can be one and the same.

Let us now discuss the typical time and velocity scales obtained in our simulations. A typical vesicle size is approximately $10$ $\mu$m, and for liposomes $\kappa\approx 10^{-19}$  Nm. Except for very curved vesicles, the typical radius of curvature $r_0$ is approximately $10$ $\mu$m as well, leading to a velocity scale of $10$ $\mu$m/s.  This gives calculated mean velocities on the order of $0.1$ $\mu$m/s for the spontaneous curvature model, and $0.5$ $\mu$m/s for the bilayer coupling model.  Translational and rotational diffusion constants for vesicles this size at room temperature are $D\approx 10^{-14}$ m$^2$/s and $D_r\approx$ $10^{-3}$ s$^{-1}$, respectively.  This implies a time scale for translational diffusion of approximately $10^4$ s, and a time scale for diffusive reorientation of approximately $10^3$ s.  Since the actuation proposed in this paper can be implemented faster than both of these time scales, significant diffusion will take place only after many actuation cycles.  For time scales much larger than $D_r^{-1}$, the effective vesicle diffusion will then be given by $D_{eff}\approx U^2/D_r$ \cite{Berg93}, which accounts for both swimming and orientation loss. The ratio $D_{eff}/D\approx 10^3$ is large, which implies that  locomotion will lead to a substantially enhanced diffusion of the vesicles over long time scales.

We have considered only two minimal models for vesicle shape change, and many possible avenues exist to  expand upon this basic model, including a study non-axisymmetric vesicles, more advanced curvature models, and arc length-dependent spontaneous curvature. Since we have assumed a quasi-static deformation, non-equilibrium effects would also have to be taken into account for fast deformations, and the shape should be fully determined as a balance between elastic and fluid forces. In addition, swimming is just one example of behavior that could be exhibited by a membrane that is actively deformed. It is perhaps the simplest transduction of geometrical deformation into mechanical work, and one that we hope provides further inspiration for the combined study of membrane physics and low Reynolds number fluid mechanics.

\section*{Acknowledgements}
This research was funded in part by the NSF (grant CBET-0746285).

\section*{Appendix: Velocity and Stress Computation}
The swimming velocity is computed at each time by solving a standard boundary integral formulation of the Stokes equations. As an application of the Lorentz reciprocal identity, the solution to Eqs.~\eqref{StokesEqs} may be written as integrations upon the surface velocity and the fluid stress, 
\begin{gather}
\b{u}(\b{x})=\frac{1}{8\pi \mu}\int_{S(t)}\b{G(x,y)}\cdot\left(\bs(\b{y})\cdot\b{\hat{n}}(\b{y})\right)\,dS_y+\frac{1}{8\pi}\int_{S(t)}\b{u(y)\cdot T(x,y)\cdot \hat{n}(y)}\,dS_y\label{StokesBIE}
\end{gather}
where
\begin{gather}
\b{G}_{ij}(\b{x,y})=\frac{\delta_{ij}}{|\b{x-y|}}+\frac{(x_i-y_i)(x_j-y_j)}{|\b{x-y}|^3},\\
\b{T}_{ijk}(\b{x,y})=-6\frac{(x_i-y_i)(x_j-y_j)(x_k-y_k)}{|\b{x-y}|^5},
\end{gather}
are the singular Stokeslet and Stresslet tensors, respectively (see Ref.~\cite{Pozrikidis92}). By introducing a complementary flow $\b{u'}$ which has the same values of the surface force $\bs\cdot \b{\hat{n}}$ as the flow $\b{u}$ on the surface $S(t)$, Eq.~\eqref{StokesBIE} may be written solely in terms of the second, double-layer integral,
\begin{gather}
\b{u(x)}=\int_{S(t)}\b{q(y)}\cdot \b{T}\b{(x,y)}\cdot \b{\hat{n}(\b{y}})\,dS_y,
\end{gather}
where $\b{q(x)}$ is an unknown density of the singular Stresslet tensor. In the limit as the $\b{x}$ approaches the body surface $S(t)$, inserting the no-slip condition for the surface velocity there, we find the expression
\begin{gather}
\b{U}+\b{u_d(x)}=\int_{S(t)}(\b{q(y)-q(x)})\cdot \b{T}\b{(x,y)}\cdot \b{\hat{n}(\b{y}})\,dS_y.
\label{DoubleLayer1}
\end{gather}
The vertical swimming velocity $U=\b{U}\cdot\b{\hat{z}}$ is related to the Stresslet density as
\begin{gather}
U=-\frac{4\pi}{A}\int_{S(t)} \b{\hat{z}}\cdot\b{q}(\b{x})\,dS\label{DoubleLayer2}
\end{gather}
(recall that $A$ is the vesicle surface area). Equation~\eqref{DoubleLayer1} is a well-posed Fredholm integral equation of the second kind for the unknown density $\b{q(x)}$, and has a unique solution. This approach is numerically better conditioned than those based on first-kind equations.

The Stresslet integral operator in Eq.~\eqref{DoubleLayer1} has a six-dimensional nullspace corresponding to rigid body motion, and in the presence of external body forces or torques this representation must be closed by a range completion technique (see Ref.~\cite{pm87}). However, in the swimming problem where the deformation velocity $\b{u_d(x)}$ is specified and there are no body forces or torques, Eqs.~(\ref{DoubleLayer1}-\ref{DoubleLayer2}) are closed and prescribe uniquely the swimming velocity $U$.

The integrand in Eq.~\eqref{DoubleLayer1} is discontinuous at the singularity but finite, so that the integrals are computed to second-order in the surface mesh element size using a standard trapezoidal quadrature (setting the quadrature weight to zero at the singularity). The axisymmetry of the problem is inserted into the definition of the body surface as well as the density $\b{q(x)}$. The number of gridpoints is chosen to be sufficiently large such that further resolution does not significantly alter the density $\b{q(x)}$ or the swimming velocity $U$.

At each time, the curve $(r(s,t),z(s,t))$ is discretized uniformly in $s$. Application of a Nystr\"om collocation method produces a linear system of equations for the density $\b{q(x)}$ at the gridpoints, which is then solved iteratively using the method GMRES \cite{ss86}, with an inversion error tolerance such that the only errors are due to discritization. Finally, the body position $z_0(t)$ is updated at each time using a second-order Runge-Kutta method. Both convergence tests and comparison with known exact solutions were used to validate the code \cite{Jeffery22,hb65,gcb66,ss96}.

Computing the hydrodynamic or swimming efficiency (which requires pointwise information about the stress $\bs$) is more difficult. Here we compute $\bs(\b{x})$ using the approach outlined below, though a more detailed description of the method and examples of its use will be featured in a subsequent paper.

Many common methods for computing the stress are developed using a first-kind boundary integral formulation of the Stokes equations, and hence can suffer from the ill-posedness of the underlying equations \cite{Pozrikidis92}. Instead, we solve for the surface stress by evaluating a hypersingular integral which may be derived from the second-kind integral equation for the velocity
(see Ref.~\cite{Pozrikidis92}),
\begin{gather}
\frac{1}{\mu}\sigma_{im}(\b{x})= \int_{S} q_j(\b{y})\b{\mathcal{L}}_{ijkm}\b{(x,y)}\hat{n}_k(\b{y})\,dS_y,\label{HyperStress}
\end{gather}
where
\begin{multline}
\b{\mathcal{L}}_{ijkm}\b{(x,y)}=-4\frac{\delta_{im} \delta_{jk}}{|\b{x-y}|^3}-6\frac{(x_k-y_k)[\delta_{jm}(x_i-y_i)+\delta_{ij}(x_m-y_m)]}{|\b{x-y}|^5}\\
-6\frac{(x_j-y_j)[\delta_{mk}(x_i-y_i)+\delta_{ik}(x_m-y_m)]}{|\b{x-y}|^5}+60\frac{(x_i-y_i)(x_j-y_j)(x_k-y_k)(x_m-y_m)}{|\b{x-y}|^7},
\end{multline}
and we have set $S(t)=S$ for clarity. The expression $\b{\mathcal{L}(x,y)}$ is achieved by differentiating the double-layer integral for the fluid velocity and including the pressure term which may also be written as an integration against $\b{q(x)}$, with $\bs=-p\b{I}+\mu\left(\nabla \b{u}+\nabla \b{u}^T\right)$ (see Ref.~\cite{Pozrikidis92}). The stress is determined on the same spatial grid as used to determine the swimming velocity, a uniform discritization in $s$ of the curve $(r(s,t),z(s,t))$ (with polar angle $\phi=0$). The integration of Eqn~\eqref{HyperStress} is performed in local polar coordinates, and the singular contributions are handled analytically as follows. The procedure follows the work of Guiggiani et al. \cite{Guiggiani98}. 

The integration of Eq.~\eqref{HyperStress} is performed on a modified surface $\tilde{S}=s_\e+(S-e_\e)$ and is taken in two parts: the portion of a sphere of radius $\e$ centered at the singular point $\b{x}$ which is internal to the body surface ($s_\e$) and intersects the surface $S$ at its boundary, and the body surface punctured by the sphere ($S-e_\e$). The modified surface limits to the body surface $S$ as $\e\rightarrow 0$. For a point $\b{x}\in S$, Eq.~\eqref{HyperStress} is written as a small $\e$ limit,
\begin{gather}
\frac{1}{\mu}\sigma_{im}(\b{x})=\lim_{\e\rightarrow 0}\Big\{\int_{S-e_\e} q_j(\b{y})\b{\mathcal{L}}_{ijkm}\b{(x,y)}\hat{n}_k(\b{y})\,dS_y+\int_{s_\e} q_j(\b{y})\b{\mathcal{L}}_{ijkm}\b{(x,y)}\hat{n}_k(\b{y})\,dS_y\Big\}\cdot\label{HyperStress2}
\end{gather}
Under the assumption that $\b{q(x)}$ is differentiable, with a derivative which is H\"older continuous, we subtract and add the density $\b{q(x)}$ and its gradient at the singular point in the second integral of Eq.~\eqref{HyperStress2},
\begin{align}
\frac{1}{\mu}\sigma_{im}(\b{x})=\lim_{\e\rightarrow 0}&\Big\{\int_{S-e_\e} q_j(\b{y})\b{\mathcal{L}}_{ijkm}\b{(x,y)}\hat{n}_k(\b{y})\,dS_y\label{HyperStress3}\\ 
&+\int_{s_\e} \Big(q_j(\b{y})-q_j(\b{x})-(x_h-y_h)q_{j,h}(\b{x})\Big)\b{\mathcal{L}}_{ijkm}\b{(x,y)}\hat{n}_k(\b{y})\,dS_y\\
&+q_{j,h}(\b{x})\int_{s_\e} (x_h-y_h)\b{\mathcal{L}}_{ijkm}\b{(x,y)}\hat{n}_k(\b{y})\,dS_y\\
&+q_j(\b{x})\int_{s_\e} \b{\mathcal{L}}_{ijkm}\b{(x,y)}\hat{n}_k(\b{y})\,dS_y,\Big\}
\end{align}
where $q_{j,h}=\partial q_j/\partial x_h$. As shown in Ref.~\cite{Guiggiani98}, the above integration may be reduced to a final formula upon the introduction of a local polar coordinate system $(\rho,\eta)$ about the target point $\b{x}(s,\phi)$, with
\begin{gather}
\phi'=\phi+\rho \cos(\eta),\,\,\,\,s'=s+\rho \sin(\eta),
\end{gather}
where $\eta\in[0,2\pi)$, $\rho\in[0,\bar{\rho}(\eta)]$, and
\begin{gather}
dS_y=J(s') ds'\,d\phi'=J(s'(\rho,\eta))\rho\,d\rho\,d\eta,
\end{gather}
with $J(s')=|\b{x}_s' \times \b{x}_\phi|$ the surface Jacobian. $\rho=\bar{\rho}(\eta)$ is the equation in the local polar coordinate system of the edge of the semi-periodic domain, $(s,\phi)\in\left([0,L]\times[0,2\pi]\right)$. The integration is assisted by the extra factor of $\rho$ in the surface area element, and the final expression for the fluid stress may be reduced to 
\begin{gather}
\frac{1}{\mu}\sigma_{im}(\b{x})=\int_0^{2\pi}\int_0^{\bar{\rho}(\eta)}\Big\{F_{ijk}(\rho,\eta)-\Big[\frac{F_{ijk}^{(-2)}(\eta)}{\rho^2}+\frac{F_{ijk}^{(-1)}(\eta)}{\rho}\Big]
\Big\}d\rho\,d\eta\\
+\int_0^{2\pi}\Big\{F_{ijk}^{(-1)}(\eta)\ln|\bar{\rho}(\eta)|-F_{ijk}^{(-2)}(\eta)\Big[\frac{1}{\bar{\rho}(\eta)}\Big]\Big\}\,d\eta,
\end{gather}
where $F_{ijk}(\b{x,y})=q_i(\b{x})\mathcal{L}_{ijk}(\b{x,y})\hat{n}_k(\b{y})$ \cite{Guiggiani98}. The functions $F_{ijk}^{(-1)}(\eta)$ and $F_{ijk}^{(-2)}(\eta)$ are the singular parts of an expansion of $F_{ijk}(\rho,\eta)$ about $\rho=0$. The integrals above all have finite integrands, and are treated using adaptive quadrature methods.

Convergence tests and comparisons with known exact solutions were used to validate the code. In particular, we have checked to ensure that the surface deformation relation of Samuel \& Stone (1996) is satisfied \cite{hb65, ss96}. With the stress $\bs$ in hand, the efficiency $\eta_H$ (Eq.~\eqref{etaH}) is determined to second-order in the grid-spacing by a simple trapezoidal quadrature. The stress need only be computed for $\phi=0$ due to axisymmetry.

As a final note, at zero Reynolds number the swimming velocity and efficiency are entirely determined by the surface deformation velocity. Other more general measures of energetic expenditure and total efficiency have been considered for other swimming systems (see Ref.~\cite{sl10}), but in this case the total efficiency will depend significantly upon the means used to produce the vesicle shape-change. In addition, should there be a fluid internal to the vesicle, for example, internal dissipation costs would be relevant in a more general measure of energetic expenditure.

\bibliography{stokesian_jellyfish.bib}

\end{document}